# Enhancing Multi-level Urban Instant Delivery Management via Infomap-based Hierarchical Community Detection

Chengbo Zhang*
Department of Urban Planning
Harbin Institute of Technology
Shenzhen 518055, China
23s156043@stu.hit.edu.cn

Lina Zhang
School of Government
Shenzhen University
Shenzhen 518055, China
2300524002@email.szu.edu.cn

Wenbin Liang
Department of Computer Science
and Engineering
Southern University of Science
and Technology
Shenzhen 518055, China
12332468@mail.sustech.edu.cn

Xiao Yang*
Faculty of Engineering
Lund University
Lund , Skane, 22100, Sweden
lxi8613ya-s@student.lu.se

*Abstract*—Efficient management of on-demand delivery systems is essential for modern urban logistics, especially in densely populated cities with complex spatial layouts. This study introduces a novel, computer-supported cooperative framework that utilizes Infomap-based hierarchical community detection to analyze spatial multilevel clustering patterns. The experiment was conducted to large scale on-demand delivery datasets from Shenzhen and Beijing, revealing integrated spatial clusters that align with cohesive urban layout. Through hierarchical detection, finer and fragmented clusters are identified, reflecting its diverse urban structure and delivery demands. The findings demonstrate the effectiveness of hierarchical community detection in uncovering spatial dependencies and optimizing resource allocation and delivery strategies. This framework provides practical insights for urban logistics, enabling tailored approaches for business hub placement, route allocation and adaptive resource management.

## I. Introduction

In the context of urban logistics, efficient management of delivery flows is increasingly critical due to the rapid growth of e-commerce and the complexity of modern cities [1], [2], [3]. The demand for fast, flexible, and cost-effective delivery systems has led to the proliferation of on-demand delivery services, where goods are transported from online retailers and warehouses to consumers in real time [4], [5]. Managing these on-demand delivery flows in large cities, with their diverse transport networks, variable service area, and fluctuating demand, presents significant challenges [6]. Traditional logistics models and route optimization algorithms often fail to capture the intricate and hierarchical spatial patterns of urban delivery systems, which are characterized by a multilevel organization of functional clusters.

A particularly complex and significant aspect of urban logistics is the optimization of multi-level city delivery clusters. These flows consist not only of the primary long-distance routes between centralized distribution hubs (such as warehouses and distribution centers) but also of local short-distance routes within neighborhoods or specific delivery zones. As urban areas become more congested and demand for instant delivery services grows, an integrated, multi-level approach to flow management is required. Additionally, most traditional models ignore the interconnectedness between regions or fail to leverage the available knowledge of community structures within the city. For example, the flow of goods in a city can vary greatly between large regional hubs and local distribution points, requiring different optimization strategies at different scales [7]. This is where community detection algorithms, especially hierarchical community detection methods, can play a crucial role in delineating both regional and local delivery networks. However, existing research rarely explore such hierarchical community structure in urban delivery system, hinders a holistic management of delivery logistics.

Community detection algorithms have been widely used in network science to analyze complex systems [8]. As freight interactions become increasingly networked [9], delivery activities can be modeled as complex spatial graphs, where nodes represent specific spatial units including distribution points and edges represent delivery flows. The weight of the edges can correspond to the order intensity. By applying community detection algorithms like Infomap, urban logistics systems can identify natural groupings of delivery points that share common flow patterns [10].

These groupings, or communities, can then be used to optimize routing and resource allocation at both the macro (regional) and micro (local) levels. For instance, at the macro level, Infomap can identify large-scale communities representing major hubs or regions within a city. These regions may be characterized by high-volume delivery routes, with multiple suppliers and demand points that interact frequently. On the micro level, Infomap can reveal smaller-scale communities of delivery points within each region, which may be optimized for last-mile delivery using more localized routes allocation. These community structures can be leveraged to guide delivery flow management at multiple levels, enabling more efficient routing, reduced congestion, and better resource utilization.

*A. Research Objective*

This paper aims to apply hierarchical community detection method to delineate multi-level urban on-demand delivery management areas, enabling smarter regional logistics policies. By constructing spatial networks from large-scale delivery datasets in Shenzhen and Beijing, the study leverages Infomap to identify hierarchical community structures, providing insights into the spatial distribution of delivery clusters and their hierarchical nesting relationships.

*B. Contribution*

This study contributes to the computer supported cooperative work on urban logistics management in two key ways:

- We propose a novel framework that combines on-demand delivery spatial network and Infomap-based hierarchical community detection for multi-level optimization, enhancing the analysis and management of urban delivery system.

- The study propose practicable insights for regional delivery strategies by computer-supported cooperative work. The findings on case cities offer practical guidance for integrating community detection algorithms into logistics frameworks to address complex urban delivery challenges, contributing to more efficient and data-driven urban logistics solutions.

## II. RELATED WORK

*A. Applications of Community Detection in Urban Mobility and Logistics*

Community detection is an essential tool for analyzing the structure of complex networks, uncovering groups or "communities" where nodes are more densely connected internally than externally. This approach has been widely applied in fields such as social networks, biological systems, and urban infrastructure to reveal hidden patterns, improve efficiency, and support decision-making in complex systems.

Recent studies have advanced the use of community detection methods to analyze urban transportation flows, revealing nested clusters that enhance city-scale logistics and transportation planning [10], [11]. Dynamic and temporal community detection techniques have been employed to capture the fluctuating rhythms of human movement, providing essential insights for adaptive logistics and responsive delivery networks [12]. Research has also focused on identifying spatial patterns and delineating regional hubs to streamline urban delivery and commuting, which contributes to more efficient resource allocation and route optimization [13], [14], [15].

However, delineating the spatial clustering structure in transportation and logistics remains largely unexplored, necessitating a hierarchical modeling framework. Table I compares various community detection algorithms for multilevel network structures. Unlike algorithms such as Louvain and Girvan-Newman, which prioritize maximizing modularity, Infomap stands out for its ability to detect hierarchical relationships and effectively handle large-scale, dynamic systems[16]. This capability makes it ideal for geo-spatial networks in urban logistics, where macro- and micro-level clustering must be simultaneously captured for multi-level optimization.

*B. Multi-Level Optimization Approaches for Delivery Flow Management*

The rapid rise of e-commerce and the increasing demand for on-demand delivery services have brought new challenges to urban logistics, particularly in last-mile delivery. Traditional optimization methods, such as the meal delivery routing problem (MDRP) [17] and its variants [18], [19], focus on optimizing individual routes but often fail to capture the complexity of spatial aggregation results from massive delivery flows. Moreover, the spatial restriction considering different scales remains unexplored. Community detection offers a complementary approach by revealing natural clusters of delivery points, enabling better resource allocation and flow management.

TABLE I. COMPARISON OF COMMUNITY DETECTION ALGORITHMS FOR MULTILEVEL NETWORK STRUCTURES

| Feature | *Infomap* | *Louvain* | *Fastgreedy* | *Label Propagation* | *Girvan-Newman* |
|---|---|---|---|---|---|
| Core Algorithm | Optimizes the map equation to compress random walk descriptions | Optimizes modularity via node movements and aggregation | Greedy optimization of modularity through hierarchical merging | Propagates labels among neighboring nodes until convergence | Sequentially removes edges based on betweenness centrality to reveal community structure |
| Computational Efficiency | High, suitable for medium-scale networks | Very efficient, suitable for large-scale networks | High, effective for large-scale networks | Very high, near-linear time complexity | High computational cost, suitable for small networks |
| Algorithm Characteristics | Effectively captures multi-level nested structures, ideal for complex networks | Efficient and scalable, performs well in large networks, but may lack fine-grained nested structure | Greedy optimization suited for networks with clear community structure | Simple and fast, but struggles with multilevel community detection | Fine-grained edge-based community discovery, well-suited for small networks |
| Suitable Scenarios | Networks requiring dynamic process capture and multilevel structure | Large-scale network community analysis with rapid computation | Networks with clearly defined community structures | Networks with dense intra-community connections | Small networks or when detailed community structure is needed |
| Multilevel Community Support | Native support through recursive nested module search | Requires iterative aggregation to form hierarchical structure | Forms hierarchical partitions through divisive clustering | Produces flat partitions without explicit multilevel support | Reveals community structure through edge removal, effective in small networks |
| | ✓✓ | ✓ | ✗ | ✗ | ✗ |

The increasing complexity of urban logistics systems requires multi-level optimization strategies that integrate macro-level planning (regional hubs) with micro-level operations (last-mile delivery) [7], [20]. However, traditional logistics models often operate in isolation, optimizing either global supply chains or local delivery paths without considering their interdependencies.

Hierarchical optimization frameworks in urban logistics have emphasized integrating regional and local operations, with Hong et al. (2019) proposing a two-tiered street network system [21] and Jia et al. (2021) introducing a framework that estimates the hierarchical mobility structure [22]. However, these studies lack adaptive spatial clustering mechanisms and fail to address on-demand delivery organization, which limits their applicability in logistic practice.

*C. Conclusion of the Related Works*

Existing work underscores the transformative potential of community detection in urban logistics and mobility but also reveals significant limitations in current approaches. While static and single-layer clustering methods provide useful insights into urban structure and flow patterns, they fall short in capturing the layered complexity of urban delivery networks, where dependencies span both macro and micro scales. Hierarchical community detection, particularly Infomap's multi-level clustering capability, offers a promising solution for bridging these scales, yet remains underutilized in logistics applications requiring dynamic and adaptive management. The key challenge is to integrate regionalization methods with real-time, multi-level optimization frameworks that accommodate shifting demands, fluctuating flow intensities, and unique urban spatial constraints. This gap suggests an opportunity: by combining hierarchical clustering with adaptive, real-time spatial optimization, urban logistics systems could be built to not only optimize delivery efficiency but also to enhance resilience and scalability, meeting the evolving demands of modern cities.

III. METHODOLOGY

This section outlines the methodology for applying Infomap-based hierarchical community detection in optimizing multi-level city on-demand delivery flow management. The goal is to use community detection to group delivery points into spatial clusters and then optimize the flow management at both regional and local levels for enhanced efficiency.

*A. Urban Delivery Network Representation*

We model the urban delivery network as a **weighted, directed graph $G = (V,E)$**, where:
- $V$ is the set of nodes, representing a grid unit.
- $E$ is the set of edges, representing delivery flows between the nodes. Each edge $e_{ij} \in E$ has a weight $w_{ij}$, representing the count of orders between $i$ and $j$.

*B. Abbreviations and Acronyms*

Infomap is a powerful algorithm for detecting community structures in large networks by minimizing the description length of a random walk on the network. The basic idea behind Infomap is to partition the network into communities such that a random walker spends more time within communities and less time moving between communities.

The key steps involved in applying Infomap to our urban logistics network are as follows:

**Graph representation:** Represent the urban delivery network as a graph $G$.

**Random walk on the network:** A random walk is initiated at each node, and the walker moves between nodes according to the probabilities defined by the edge weights. The transition probability $P_{ij}$ between nodes $i$ and $j$ is computed as follows:

$$P_{ij} = \frac{w_{ij}}{\sum_{k \in N(i)} w_{ik}}, \quad \forall j \in N(i) \qquad (1)$$

where $N(i)$ is the set of neighboring nodes of node $i$, and $w_{ij}$ is the weight of the edge between nodes $i$ and $j$. This transition probability reflects the likelihood of the random walker moving from node $i$ to node $j$.

**Community detection:** Infomap detects community structures by minimizing the total description length of the random walk on the network. The description length consists of two components:

- Internal Description: The cost of describing the steps taken within a community.
- External Description: The cost of describing the transitions between different communities.

Infomap seeks to partition the network $G$ into $m$ modules by minimizing the *expected description length L(M)*, defined as:

$$L(M) = q\ H(\zeta) + \sum_{i=1}^{m} p_i\ H(P_i) \qquad (2)$$

where: $q$ is the robability of a random walker transitioning between modules. $H(\zeta)$ is the entropy of module encoding (prefix codes). $H(P_i)$ is the entropy of movements within module $i$ (suffix codes). $p_i$ is the probability of accessing module $i$, including the likelihood of internal node visits and exits.

This formulation incorporates both the *entropy of inter-module movements* and the *entropy of intra-module movements*, ensuring a holistic representation of the network's structure.

*C. Algorithmic Implementation*

Once primary communities are identified, Infomap recursively applies the same optimization process within each community to identify sub-communities. This hierarchical iteration continues until no further improvement in the description length can be achieved at finer partitions. The hierarchical structure emerges naturally from this process as larger macro-level communities are subdivided into increasingly localized micro-level communities.

- **Initial partitioning:** A deterministic greedy search algorithm iteratively minimizes $L(M)$, producing an optimal two-level division of the network into communities. The modules reflect significant spatial clusters within the urban delivery network.

- **Recursive subdivision:** Infomap extends the two-level description to a multilevel framework by recursively applying the same optimization process within each detected community. This hierarchical subdivision continues until no further improvement in *L(M)* is achievable, uncovering nested community structures at various levels of granularity.

- **Hierarchical features**: The resulting hierarchy reveals both macro-level communities (regional clusters) and micro-level communities (localized zones), reflecting the topological and functional relationships within the network. This multilevel perspective provides a comprehensive understanding of the network's modularity and flow dynamics.

- **Community assignment:** After applying Infomap, each node (spatial units) is assigned to one or more communities. These assignments represent natural groupings of delivery points that can be used for optimizing delivery flows at different scales.

## IV. EXPERIMENTS AND RESULTS

To examine the spatial clustering structure of on-demand delivery networks, we applied Infomap-based hierarchical community detection in two major cities, Shenzhen and Beijing.

### A. Data description

Fig. 1 illustrates the delivery networks built in Shenzhen and Beijing.

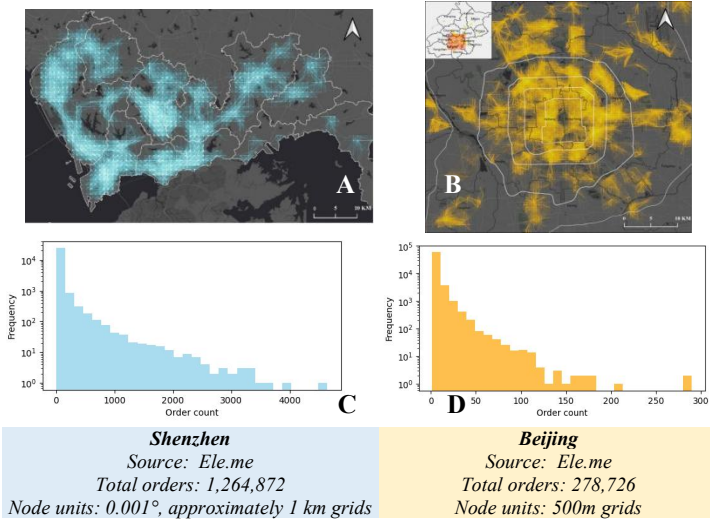

*Shenzhen*
Source: Ele.me
Total orders: 1,264,872
Node units: 0.001°, approximately 1 km grids

*Beijing*
Source: Ele.me
Total orders: 278,726
Node units: 500m grids

Fig. 1. Spatial distribution and order frequency of delivery networks in Shenzhen and Beijing. (A) Delivery network in Shenzhen (B) Delivery network in Beijing (C) Frequency distribution of order counts in Shenzhen (D) Frequency distribution of order counts in Beijing

**Shenzhen:** The Shenzhen dataset, RL-Dispatch, is sourced from the Ele.me food delivery platform and provided by Alibaba [23]. Spanning one month in 2019, this dataset provides detailed information on delivery activities in Shenzhen, including sender and receiver locations aggregated to a resolution of 0.001 degrees (approximately 1 km). These points are used to construct Voronoi polygons, which serve as spatial nodes for further analysis.

**Beijing:** The Beijing dataset, also sourced from the Ele.me platform, documents delivery orders completed in February 2020. It includes comprehensive spatial and temporal details for each transaction, such as pickup and delivery locations, order creation time, and delivery completion time. For this dataset, spatial nodes are created by dividing the city into 500-meter grid units, enabling fine-grained spatial analysis.

### B. Hierarchical Community Detection results

In this section, we present the results of the hierarchical community detection applied to our on-demand delivery network model, implemented in a Python environment using the Infomap package. To determine the community divisions at various hierarchical levels, we adjusted the *depth_level* parameter, which controls the granularity of community detection, with higher values corresponding to more refined divisions of the network.

Our results show that, for both the Shenzhen and Beijing datasets, the *depth_level* parameter reached stability at a value of 3, indicating that the optimal community structure for both cities is defined by three hierarchical levels. This suggests that, at this depth, the delivery network in both cities can be effectively partitioned into three distinct layers of communities, which represent different scales of the on-demand delivery system.

Fig. 2 shows the hierarchical delivery communities in Shenzhen, specifically:

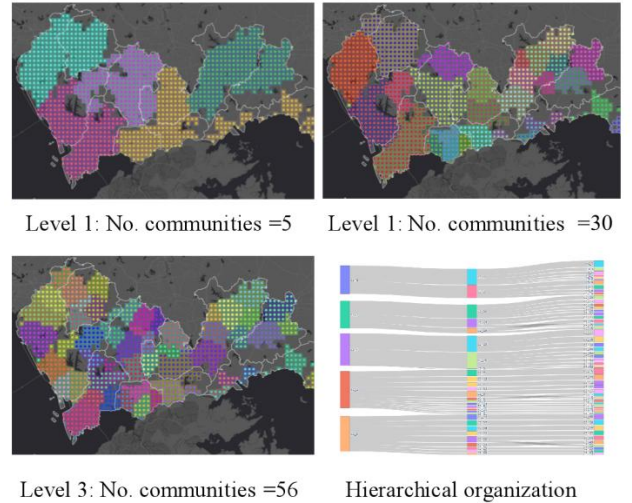

Level 1: No. communities = 5    Level 1: No. communities = 30

Level 3: No. communities = 56    Hierarchical organization

Fig. 2. Hierarchical delivery communities in Shenzhen

**Level 1:** The network was divided into 5 primary communities, representing large regional clusters. These broad spatial groupings highlight major delivery flow regions across Shenzhen, providing a high-level overview of the city's delivery dynamics.

**Level 2:** At this intermediate level, 30 communities were detected, representing more granular clusters. These communities correspond to specific urban sub-regions, likely corresponding to different neighborhoods or delivery zones, each with distinct patterns of demand and flow.

**Level 3:** The finest level identified 56 communities, each corresponding to even smaller clusters within Shenzhen. These communities represent highly localized delivery zones, capturing the micro-level spatial structures where delivery demand is concentrated.

These results illustrate a clear hierarchical structure of delivery communities. At Level 1, large regional clusters break down into smaller, more localized communities at Levels 2 and 3, providing a layered understanding of the delivery network. The Sankey plot for Shenzhen's community structure visually represents this hierarchy, showing how each community at a finer level is nested within a broader cluster at higher levels, providing a detailed view of the network's organization.

Similarly, Fig. 3 shows the hierarchical delivery communities in Beijing, specifically:

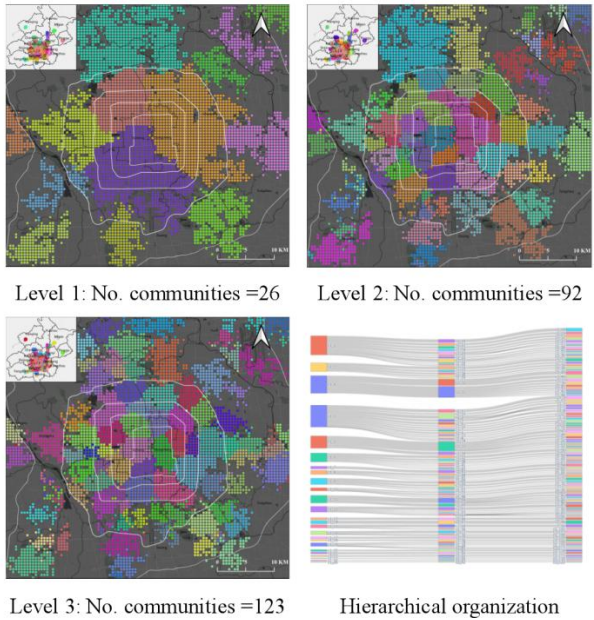

Fig. 3. Hierarchical delivery communities in Beijing

**Level 1:** The network was divided into 26 communities, representing large spatial clusters across the city. Each community at this level covers a broad urban area, with an average of 223 grid units per community, indicating substantial spatial coverage for each primary community.

**Level 2:** At this intermediate level, 92 communities were detected, with an average of 63 grid units per community. These mid-sized communities likely correspond to districts or zones with higher delivery demand within the larger areas defined at Level 1.

**Level 3:** The finest level identified 123 communities, with each community covering an average of 47 grid units. This level corresponds to neighborhood-scale clusters, capturing the detailed spatial structure of delivery patterns in more localized areas, with each community covering an average area of approximately 11.75 km² .

These hierarchical clusters in Beijing effectively capture the layered nature of delivery flows, with broader regional patterns at Level 1 gradually subdividing into smaller, more localized zones at Levels 2 and 3. The Sankey plot for Beijing's community structure visually represents this nested hierarchy, showing how the smallest clusters at Level 3 are nested within larger clusters at Levels 2 and 1. This visualization underscores the spatial organization of delivery flows, revealing how smaller, localized delivery zones contribute to larger regional patterns.

Fig. 4 illustrates the distribution of community size at three hierarchical levels for the cities of Shenzhen and Beijing. The results reveal notable differences in the spatial clustering structures of these cities, reflecting the distinct urban layouts and delivery needs in each location.

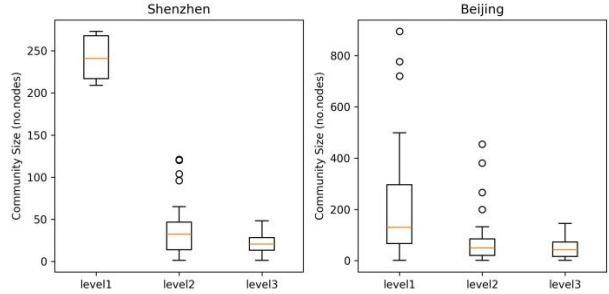

Fig. 4. Distribution of detected community size across hierarchical levels

## V. CONCLUSION

This study applied Infomap-based hierarchical community detection to analyze urban on-demand delivery networks in Shenzhen and Beijing. The results reveal significant spatial clustering patterns, with Shenzhen showing broad, integrated clusters reflective of its cohesive urban structure and Beijing exhibiting fine-grained, fragmented clusters corresponding to its complex urban layout and diverse delivery needs. These findings underscore the utility of hierarchical community detection in identifying multi-level delivery flow structures and optimizing logistics strategies tailored to specific urban contexts.

The research highlights the scalability and adaptability of Infomap in capturing nested community structures, providing a robust framework for multi-level urban delivery management (Fig. 5). By uncovering spatial hierarchies, this approach enables data-driven decision-making in hub placement, resource allocation, and route optimization, offering a practical foundation for improving delivery efficiency in densely populated cities.

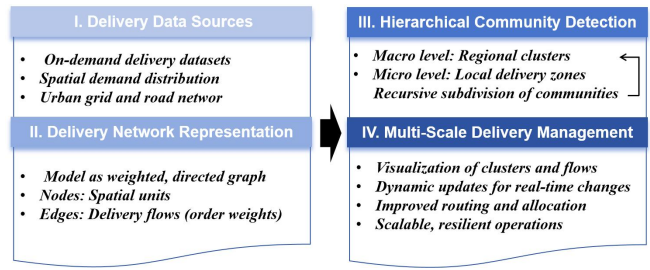

Fig. 5. CSCW framework for multi-level urban delivery management

Future research should explore the integration of dynamic temporal data to account for fluctuations in delivery demand and traffic conditions, enhancing the adaptability of community detection models. Additionally, incorporating external variables, such as socioeconomic and land use data, could provide deeper insights into the drivers of delivery patterns [24]. Expanding the framework to a broader range of cities with varying urban characteristics would help validate the generalizability of the findings. Addressing these avenues will contribute to more comprehensive and sustainable solutions for urban logistics challenges, fostering smarter and more efficient city-wide delivery systems.